# Phase Separation and the Low-Field Bulk Magnetic Properties of $Pr_{0.7}Ca_{0.3}MnO_3$

I. G. Deac[1,3], J. F. Mitchell[2], and P. Schiffer[1,*]

[1] *Department of Physics, Pennsylvania State University, University Park, PA 16802*

[2]*Material Science Division, Argonne National Laboratory, Argonne, IL 60439*

[3]*Department of Physics, Babes-Bolyai University, 3400 Cluj-Napoca, Romania*

## Abstract

We present a detailed magnetic study of the perovskite manganite $Pr_{0.7}Ca_{0.3}MnO_3$ at low temperatures including magnetization and a.c. susceptibility measurements. The data appear to exclude a conventional spin glass phase at low fields, suggesting instead the presence of correlated ferromagnetic clusters embedded in a charge-ordered matrix. We examine the growth of the ferromagnetic clusters with increasing magnetic field as they expand to occupy almost the entire sample at H ~ 0.5 T. Since this is well below the field required to induce a metallic state, our results point to the existence of a field-induced ferromagnetic insulating state in this material.

---

[*]Corresponding author: schiffer@phys.psu.edu

## I. INTRODUCTION

There has been much recent interest in the relation between the structural, magnetic and transport properties of perovskite manganese oxides with the general formula $Ln_{1-x}A_xMnO_3$, where Ln is a lanthanide and A is an alkaline earth. These materials display a number of remarkable properties including an anomalously large negative magnetoresistance, the so-called "colossal" magnetoresistance (CMR) effect.[1] The fascinating physics of these materials is driven by a close coupling between lattice, electronic, and magnetic degrees of freedom, which has recently been shown to commonly result in electronic phase separation between different magnetoelectronic states at low temperatures.[2]

One series of these oxides, $Pr_{1-x}Ca_xMnO_3$, is insulating for any Ca doping in zero magnetic field.[3,4,5,6,7,8,9,10,11,12] although the energy difference between the insulating state and a metallic phase for $x = 0.25$-$0.50$ is expected to be unusually small.[13] For $x \sim 0.3$, Hwang et al.[14] and Tomioka et al.[6] showed that the insulating phase can be driven metallic at low temperatures by applying a magnetic field of a few tesla. Additionally, a metallic phase can be induced by the application of light,[15] pressure,[10] x-ray irradiation,[11] or a high electric field,[15] making this composition perhaps the most interesting of the perovskite manganites. The nature of the insulating magnetic ground state of this material in small magnetic fields is, however, particularly interesting. The first x-ray and neutron powder diffraction studies reported a charge-ordering transition in the range 200 - 250 K, followed by antiferromagnetic and ferromagnetic transitions at about 130 K and 110 K, respectively.[3] Jirak et al. then proposed the coexistence of two magnetic phases in equal quantities at low temperatures.[4,5] The neutron study of Yoshizawa et al. suggested a canted antiferromagnetic transition at 110 K, and, because of the small magnetic moment and a diffuse scattering peak, they suggested that some fraction of the moments formed a spin-glass-like state.[7,8] Cox et al. observed the development of a ferromagnetic component at about 120 K, and they interpreted this in terms of ferromagnetic clusters with an associated lattice distortion from the average structure, in an inhomogeneous system.[10] The neutron scattering study of Radaelli et al.[16] concluded that a microscopically inhomogeneous state develops with non-metallic ferromagnetic clusters interspersed in a charge-ordered antiferromagnetic matrix. Further neutron scattering studies by Katano et al. also suggested the occurrence of the ferromagnetic clusters which grow in a magnetic field to become part of a homogenous long-range ferromagnet.[17] Frontera et al. also investigated the coexistence of ferromagnetic metallic and antiferromagnetic charge-ordered states through zero-field muon spin relaxation, neutron diffraction, calorimetric and magnetic measurements.[18] They ruled out a spin reorientation from a pure antiferromagnetic structure to a canted magnetic ordering, and they suggested that ferromagnetic and antiferromagnetic regions have a spatial distribution strongly interpenetrated with variable size clusters of the minority phase densely scattered within the majority phase. Very recent magnetocaloric measurements show enhanced heat release at low fields after zero-field cooling,[19] consistent with the irreversibility associated with spin-glass-like properties.[20]

We report a detailed study of the bulk magnetic properties of $Pr_{0.7}Ca_{0.3}MnO_3$ at low temperatures. Our magnetization and a.c. magnetic susceptibility measurements are not consistent with a spin glass phase at low fields and lend further support to models of the coexistence of ferromagnetic clusters in the charge-ordered state at low magnetic fields. Our studies allow us to track the development of the clusters with increasing magnetic field to form a nearly fully ferromagnetic insulating state for $H \gtrsim 0.5$ T which then undergoes an irreversible transition to a metallic ferromagnetic state at $H \sim 4$ T.

**II. EXPERIMENTAL DETAILS**

We studied both a single crystal of $Pr_{0.7}Ca_{0.3}MnO_3$ grown in a floating zone image furnace and a ceramic sample synthesized by a standard solid state technique. Both samples show qualitatively similar behavior, and the data presented here are for the single crystal sample.[21] Zero-field-cooled (ZFC) and field-cooled (FC) magnetization was measured at various applied fields, from 0.0005 to 7 T in a Quantum Design SQUID magnetometer in the temperature range from 4 to 250 K. A Quantum Design PPMS cryostat was used for a.c. susceptibility measurements in magnetic fields up to 4 T and temperatures ranging from 4 to 300 K. The amplitude of the a.c. field was 0.001 T, although the results using lower driving fields ($H_{ac} = 10^{-4}$ T) are not qualitatively different.

## III. RESULTS

Figure 1 depicts the temperature dependence of the FC and ZFC magnetization. The magnetization has a history dependence with a bifurcation between ZFC and FC data at an irreversibility temperature, $T_{irr}$.[22] Note that the qualitative behavior of the magnetization changes with field, i.e. the sharp maximum in $M_{ZFC}$ seen at low fields broadens in higher fields and shifts to lower temperatures for H > 0.03 T. For the field range from 0.7 T to 2 T, $M_{ZFC}$ increases monotonically without reaching a maximum, but a new broad maximum appears at much lower temperatures at the highest fields.

The temperature dependence of the real part of a.c. susceptibility at 50 Hz is shown in the upper panel of figure 2 for fields up to 4 T. We find that $\chi'(T)$ displays a rather sharp maximum at $T \cong 106$ K in low fields, and that the peak shifts to lower temperatures at lower frequencies (see inset). The out-of-phase component of the complex susceptibility, $\chi''$, is shown in the lower panel of figure 2. We find that $\chi''$ also has a peak at $T \cong 104$ K with a smaller local maximum at T ~ 110 K suppressed in fields higher than 0.01 T.[23] As shown in the inset, the amplitude of the $\chi''$ peak at $H = 0$ increases with decreasing frequency while the corresponding peak temperature decreases.

## IV. DISCUSSION

As discussed above, different neutron diffraction studies have suggested magnetic ground state in low fields might be either homogeneous spin-glass-like or phase separated on a large scale, and the qualitative features of our data do not differentiate between these possibilities. The peak in the a.c. susceptibility and differences between the field-cooled and zero-field-cooled magnetization are characteristic of both spin glasses[24,25,26] and inhomogeneous clustered systems.[27,28,29,30] The frequency dependence of the cusp in $\chi'(T)$ is also qualitatively consistent with either a spin glass or a cluster system, but it can be quantified through the frequency dependence of the freezing temperature as given by the peak position ($T_f$). As shown in figure 3, we find that $T_f$ is linear in the logarithm of the frequency with a normalized slope $p = \Delta T_f/T_f \Delta log\omega$. We find that $p = 0.00154$, which is much lower than typical values for canonical spin glass systems in which $p$ ranges from 0.0045 to 0.28.[24] We attempted to fit these data to Arrhenius behavior, i.e. $\omega = \omega_0 e^{-E_a/k_B T_f}$, which is expected for a superparamagnet, but the fit yielded an unphysically large value of $\omega_0 > 10^{500}$ Hz. A fit to the dynamic scaling relation, $\omega = \omega_0 (T/T_c - 1)^{z\nu}$ also yielded unphysical values ($\omega_0 \sim 10^{30}$ Hz, $z\nu \sim 14$). The data could be fit to a Vogel-Fulcher law ($\omega = \omega_0 e^{-E_a/k_B(T_f - T_o)}$) which presumes correlations between spin clusters, but the uncertainty in the fit parameters was too large to provide physical insight. We can thus conclude from the frequency dependence only that the system is not a simple superparamagnet and also probably not a conventional spin glass.

Our measurements of $\chi''$ give stronger evidence that the low field ground state does not behave like a conventional spin glass. The peak in $\chi''$ increases with decreasing frequency as shown in the inset to lower panel of figure 2. This is qualitatively different from the behavior of most spin glasses in which we expect an increase of the peak magnitude as frequency is increased.[24,31] Furthermore, the relative positions of the $\chi''$ peak and $\chi'$ peak are not typical for a spin glass material in that the sharp rise in $\chi''(T)$ is not at the same temperature as the sharp peak in $\chi'(T)$.[24] Our data thus appear to support the most recent neutron diffraction studies,[10,16,17,18] suggesting the coexistence of a charge-ordered phase and clusters of a ferromagnetic phase with a ferromagnetic $T_c \sim 110$ K.

We can analyze the magnetization data in a variety of ways in order to better understand the nature of the proposed ferromagnetic clusters. We plot the field dependence of the analyzed data in figure 4, showing the low temperature magnetization in figure 4a; the irreversibility temperature ($T_{irr}$) in figure 4b;[22] the low temperature relative difference between the field-cooled and zero-field-cooled magnetization, $\Delta M = (M_{ZFC} - M_{FC})/M_{FC}$, in figure 4c; and the difference between $T_{irr}$ and the temperature of the maximum in the ZFC magnetization ($T_{max}$) in figure 4d. We include a field-temperature phase diagram based on previous transport measurments[6] as an inset for comparison.

The data seem to support the existence of four different regimes of behavior with increasing magnetic field. For $H \lesssim 0.02$ T, the ferromagnetic moment is strongly dependent on the magnetic history in that the difference between $M_{ZFC}$ and $M_{FC}$ is relatively large. Additionally, in this regime we find that $\Delta T = T_{irr} - T_{max}$ is small and relatively constant. The strong history dependence suggests that either the intrinsic anisotropy of the clusters or the inter-cluster interactions dominate the effects of the external magnetic field, and the relatively small $\Delta T$ indicates that the ferromagnetic clusters at such low fields are relatively uniform in size and in blocking field.[25,32] The constant nature of $\Delta T$ with increasing field suggests that the applied field does not change the size of the ferromagnetic regions in this field range, but that the increasing magnetization comes from alignment of their moments with the increasing field.

For $0.02 \lesssim H \lesssim 0.5$ T, $T_{irr}$ decreases slightly, but $\Delta T$ increases sharply and $\Delta M$ decreases sharply, while the total magnetization rises to nearly the full ferromagnetic value. The increase in $\Delta T$ presumably corresponds to an increasing size distribution among the clusters,[25,32] indicating that they are growing in size in this regime rather than simply reorienting their moments. We associate the decrease in $\Delta M$ with the increasing Zeeman energy relative to the anisotropy energy of the clusters, possibly due to the clusters becoming more spherical.

At $H \sim 0.5$ T, we see that $T_{irr}$ sharply decreases as $\Delta T$ and $M$ saturate, and $\Delta M$ approaches zero. Given the nearly saturated value of the total magnetization, it appears that the ferromagnetic phase fills nearly the entire volume of the sample at this field even though transport measurements in this regime do not show the material to be conducting.[6,33] As can be seen from the $M(H)$ data, the material does not have a full ferromagnetic magnetization in this regime, either due to canting or a residual presence of the charge-ordered antiferromagnetic phase. We hypothesize that the insulating phase above 0.5 T could correspond to the proposed "reverse orbital ordering" of reference 16, which is ferromagnetic but not conducting. At $H \sim 4$ T, the field at which there is the insulator-metal transition[6] and an associated steplike increase in $M(H)$ measured after zero-field-cooling,[19] we see that $T_{irr}$ drops sharply again. We presume that this feature corresponds to the sample undergoing the insulator to metal transition although the large fluctuations in the electrical transport properties of our sample[15,34] preclude a definitive proof of this hypothesis.

The growth of the ferromagnetic clusters can also be seen in the a.c. susceptibility data of figure 2. As the static magnetic field is increased, the peak in $\chi'$ broadens and the temperature of the maximum in susceptibility shifts to lower temperature. At fields above $H \sim 0.2$ T a new maximum appears at $T \sim 120$ K which corresponds to the approximate ferromagnetic transition temperature $T_c$ as indicated by an inflection in $M(T)$.[35] This pair of maxima has been observed in other ferromagnetic materials in which there are competing interactions,[26] and they can possibly be attributed to critical fluctuations near the transition to ferromagnetism and blocking of the domain orientations at lower temperatures (the lower temperature peak does seem to approximately track $T_{irr}$ below $H = 0.5$ T). We find that $\chi''(T)$ also shows a small maximum at $\sim 12$ K, which increases in magnitude with increasing frequency and is suppressed in magnetic fields above $H \sim 0.5$ T. This peak was found to be enhanced for the ceramic sample at 14.6 K, and we hypothesize that it could be related to the blocking of isolated spins between ferromagnetic clusters as has been suggested for similar features in traditional spin glass materials.[36] This explanation is consistent with the peak's vanishing at the same field where the magnetization reaches nearly full saturation.

## V. CONCLUSIONS

The magnetic behavior of $Pr_{0.7}Ca_{0.3}MnO_3$ at low fields displays some features that are qualitatively consistent with spin-glass-like or superparamagnetic behavior, but our a.c. susceptibility data are not fully consistent with either. We conclude that the low field data can best be explained as resulting from an inhomogeneous phase of ferromagnetic clusters within a non-ferromagnetic matrix, in agreement with previous neutron scattering measurements. These clusters expand with increasing magnetic field to include the entire volume of the sample within a ferromagnetic insulating state, and then the sample undergoes a first order insulator-metal transition at higher fields.

We gratefully acknowledge financial support from NSF grant DMR 97-01548, the Alfred P. Sloan Foundation and the Dept. of Energy, Basic Energy Sciences-Materials under contract No.W-31-109-ENG-38.

**Figure captions:**

Figure 1. Field cooled (open symbols) and zero field cooled (closed symbols) magnetization of $Pr_{0.7}Ca_{0.3}MnO_3$ as a function of temperature, measured in the field ranges (a) 0.005 - 0.07 T; (b) 0.1- 0.5 T; (c) 1 – 7 T.

Figure 2. Real ($\chi'$) and imaginary ($\chi''$) components of the ac susceptibility of $Pr_{0.7}Ca_{0.3}MnO_3$ measured in applied static magnetic fields. The insets indicate the variation of $\chi'$ respectively $\chi''$ with frequency.

Figure 3. The variation of the $\chi'$ peak temperature with the frequency of the driving a.c. field in the range 10 – 10000 Hz

Figure 4. The field dependence of the analyzed magnetization data. a.) $M_{ZFC}$ and $M_{FC}$ at 4 K. b.) The irreversibility temperature, $T_{irr}$[22] (the inset shows the phase diagram obtained previously by electrical measurements,[6] with the regime of history dependence shown by the hatched area). c.) The relative history dependence of the magnetization $\Delta M = (M_{FC}-M_{ZFC})/M_{ZFC}$ at 4 K. d.) The difference between the irreversibility temperature and the temperature of the maximum in the ZFC magnetization, $\Delta T = T_{irr} - T_{max}$.

**REFERENCES**

1. For reviews see: A. P. Ramirez, J. Phys. Condens. Matter **9**, 8171 (1997) and J. M. D. Coey, M. Viret, and S. von Molnar, Adv. Physics **48**,167 (1999).

2. A. Moreo, S. Yunoki, E. Dagotto, Science **283**, 2034-2040 (1999); D. Khomskii, Physica B **280**, 325-330 (2000) and references therein.

3. E. Pollert, S. Krupicka, and E. Kuzmicova, J. Phys. Chem. Solids **43**, 1137 (1982).

4. Z. Jirak, S. Krupicka, V. Nekvasil, E. Pollert, G. Villeneuve, and F. Zounova, J. Magn. Magn. Mater. **15-18**, 519 (1980).

5. Z. Jirak, S. Krupicka, Z. Simsa, M. Dlouha and S. Vratislav, J. Magn. Magn. Mater. **53**, 153 (1985).

6. Y. Tomioka, A. Asamistsu, Y. Moritomo and Y. Tokura, J. Phys. Soc. Jpn. **64**, 3626 (1995). Y. Tomioka, A. Asamistsu, H. Kuwahara and Y. Moritomo, Phys. Rev. B **53**, R1689 (1996).

7. H. Yoshizawa, H. Kawano, Y. Tomioka, Y. Tokura, J. Phys. Soc. Jpn. **65**, 1043 (1996).

8. H. Yoshizawa, H. Kawano, Y. Tomioka, Y. Tokura, Phys. Rev. B **52**, R13145 (1995).

9. Y. Moritomo, H. Kuwahara, Y. Tomioka and Y. Tokura, Phys. Rev. B **55**, 7549 (1997).

10. D. E. Cox, P. G. Radaelli, M. Marazio, S-W. Cheong, Phys. Rev. B **57**, 3305 (1998).

11. K. Miyano, T. Tanaka, Y. Tomioka, and Y. Tokura, Phys. Rev. Lett. **78**, 4257 (1997).

12. V. N. Smolyaninova, Amlan Biswas, X. Zhang, K. H. Kim, Bog-Gi Kim, S-W. Cheong, and R. L. Greene Phys. Rev. B **62**, 6093 (2000).

13. T. Hotta and E. Dagotto, Phys. Rev. B **61**, 11879 (2000).

14. H. Y. Hwang, S-W. Cheong, P. G. Radaelli, M. Marezio, and B. Batlogg, Phys. Rev. Lett. **75**, 914 (1995).

15. A. Asamistsu, Y. Tomioka, H. Kuwahara and Y. Tokura, Nature (London) **388**, 50 (1997); J. Stankiewicz, J. Sese, J. Garcia, J. Blasco, and C. Rillo, Phys. Rev. B **61**, 11236 (2000).

16. P. G. Radaelli, G. Innanone, D, E. Cox, M. Marazio, H. Y. Hwang and S-W. Cheong, Physica B **241-243**, 295 (1998). P. G. Radaelli, R. M. Ibberson, D. N. Argyriou, H. Casalta, K. H. Andersen, S-W. Cheong and J. F. Mitchell., unpublished (cond-mat/0006190).

17. S. Katano, J. A. Fernandez-Baca, andY. Yamada, Physica B **276-278**, 787 (2000).

18. C. Frontera, J. L. Garcia-Munoz, A. Llobet, J. S. Lord, A. Planes, Phys. Rev. B **62**, 3381 (2000).

19. M. Roy, J. F. Mitchell, A. P. Ramirez, and P. Schiffer, Phys. Rev. B **62**, 13876 (2000) and Phil. Mag. B (in press).

20. Y. K. Tsui, C. A. Burns, J. Snyder, and P. Schiffer, Phys. Rev. Lett. **82** 3532 (1999) and references therein.

21. These samples are the same which were studied in previous thermodynamic experiments.[19]

22. The irreversibility temperature was taken as the point where *($M_{FC}$-$M_{ZFC}$)/$M_{ZFC}$* and *[∂($M_{FC}$-$M_{ZFC}$)/∂T]/[∂$M_{ZFC}$/∂T]* deviated below the noise (which was always less than 0.1%).

23. We hypothesize that the double peak in $\chi''$ at low fields is associated with domain formation within the ferromagnetic clusters.

24. J. A. Mydosh, *Spin Glasses: An Experimental Introduction* (Taylor & Francis, London, 1993).

25. S. Chikuzami, *Physics of Ferromagnetism* (Clarendon, Oxford, 1997).

26. G. Williams, in *Magnetic Susceptibility of Superconductors and Other Spin Systems* ed. R. A. Hein, T. L. Francavilla and D. H. Liebenberg (Plenum, New York, London, 1991) p.475 and references therein.

27. J. L. Tholence, in *Magnetic Susceptibility of Superconductors and Other Spin Systems* ed. R. A. Hein, T. L. Francavilla and D. H. Liebenberg (Plenum, New York, London, 1991) p.503.

28. J. L. Dormann, R. Chrkaoui, L. Spinu, M. Nogues, F. Lucari, F. D'Orazio, D. Fiorani, A. Garcia, E. Tronc, J. P. Jolivet, J. Magn. Magn. Mater. **187**, L139 (1985).

29. H. Mamiya, I. Nakatani and T. Furubayashi, Phys. Rev. Lett. **80**, 177 (1998).

30. J. A. De Toro, M. A. Lopez de la Torre, J. M. Riveiro, R. Saez Puche, A. Gomez-Herrero and L. C. Otero-Diaz, Phys. Rev. B **60**, 12918 (1999).

31. D. S. Fischer, Phys. Stat. Sol. **130,** 13, (1985).

32. L.C.C.M. Nagamine, B. Mevel, B. Dienty, B. Rodmacq, J.R. Regnard, C. Revenant-Brizard, and I. Manzini, J. Magn. Magn. Mater. **195**, 437 (1999).

33. M. Roy, Ph.D Thesis, 1999 (University of Notre Dame, unpublished).

34. A. Anane , J. P Renard, L. Reversat, C. Dupas, P. Veillet, M. Viret, L. Pinsard, A. Revcolevschi, Phys. Rev. B **59**, 77 (1999).

35. The ferromagnetic $T_c$ of our sample, as estimated from the common inflection points of both $M_{ZFC}$ and $M_{FC}$, is nearly constant of about 112 K for low fields (up to 0.4 T) and then increases approximately linear at 5.9 K/T.

36. D. Hüser, L. E. Wenger, A. J. van Duynevelt, and J. A. Mydosh, Phys. Rev. B **27** 3100 (1983).

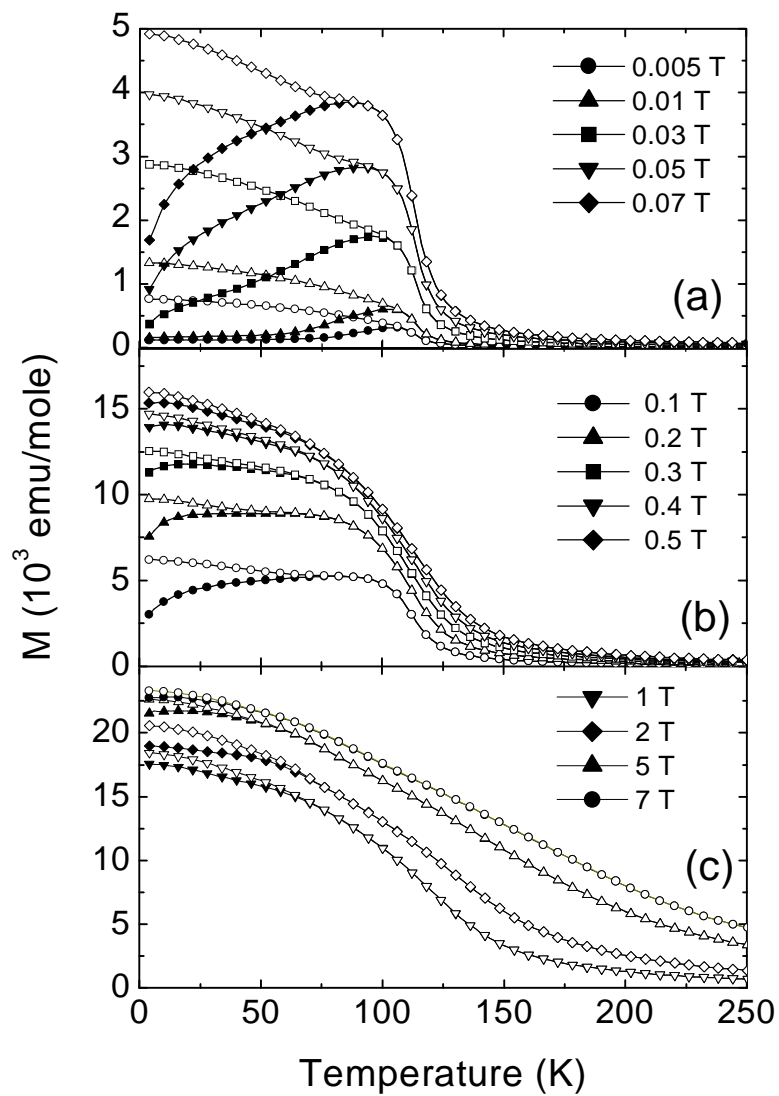

Figure 1 Deac et al.

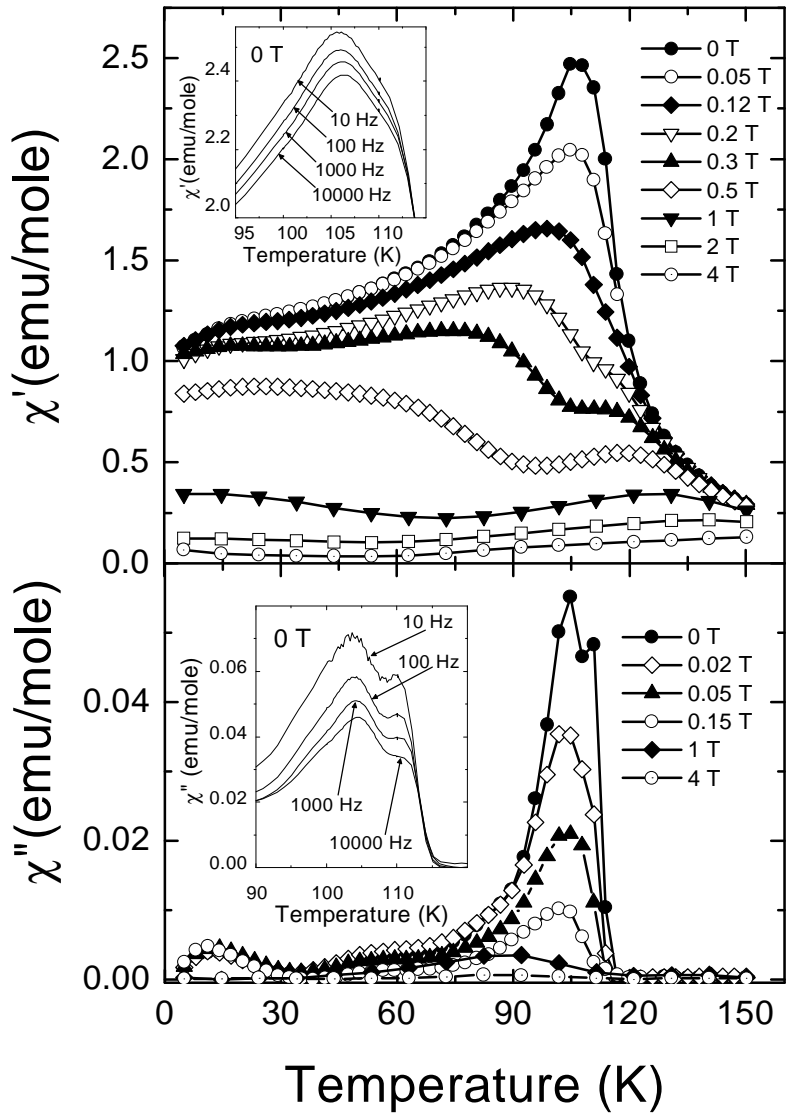

Figure 2 Deac et al.

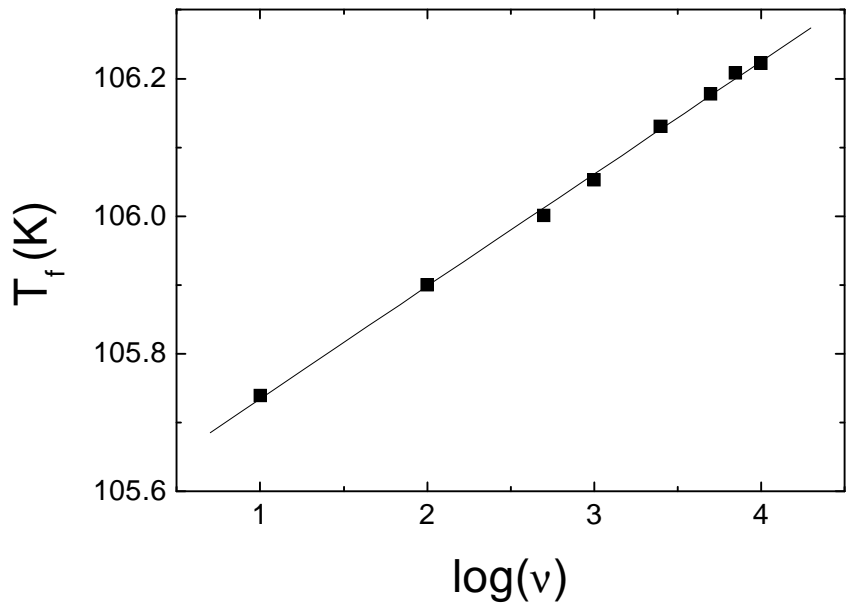

Figure 3  Deac et al.

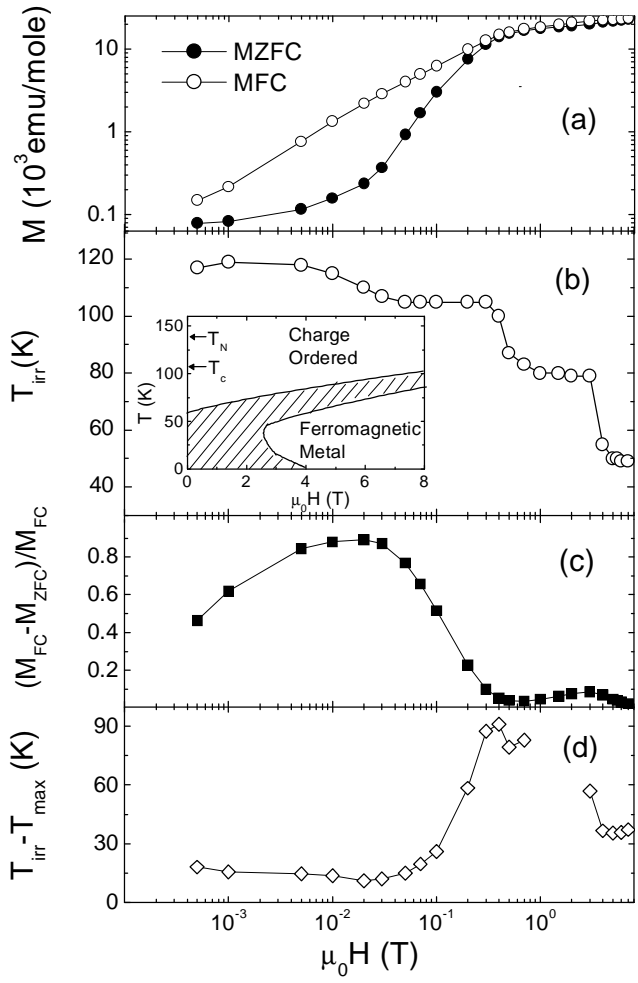

Figure 4 Deac et al.